\documentclass[lettersize,journal]{IEEEtran}
\IEEEoverridecommandlockouts

\usepackage{amsmath,amsfonts}
\usepackage{algorithm}
\usepackage{array}
\usepackage[caption=false,font=normalsize,labelfont=sf,textfont=sf]{subfig}
\usepackage{stfloats}
\usepackage{url}
\usepackage{verbatim}
\usepackage{graphicx}
\usepackage{multicol}
\usepackage{lipsum}
\usepackage{caption}
\usepackage{hyperref}
\usepackage{booktabs}
\usepackage{tikz}
\usetikzlibrary{positioning,arrows.meta}
\usepackage{titlesec}
\usepackage{booktabs,array}

\usepackage[framemethod=TikZ]{mdframed}
\usepackage{xcolor} 
\definecolor{quote}{HTML}{9673A6}

\newcommand{\ta}[1]{\textcolor[HTML]{0D0D0D}{\textbf{#1} }}

\definecolor{customcolor}{HTML}{747474}

\newenvironment{fancy}[2][]{
    \mdfsetup{
        skipabove=1pt, 
        innerlinewidth=1pt, innerlinecolor=#2, 
        linewidth=0pt,
        backgroundcolor=#2!20 
    }
    \begin{mdframed}}
    {\end{mdframed}}

\newcommand{\cristysbox}[1]{

\begin{fancy}[]{customcolor} 
\noindent\parbox{0.96\linewidth}
{\vspace{1px}
#1
\vspace{5px}} 
\vspace{-5pt}
\end{fancy}
}

\renewcommand\theparagraph{\alph{paragraph}} 

\titleformat{\paragraph}[runin]{\normalfont\normalsize\bfseries}{(\theparagraph)}{0.5em}{}

\usepackage{multirow}
\usepackage{cite}
\usepackage{amsmath,amssymb,amsfonts}
\usepackage{algorithmic}
\usepackage{graphicx}
\usepackage{textcomp}
\usepackage{longtable}
\usepackage{enumitem}
\usepackage{xcolor}
\def\BibTeX{{\rm B\kern-.05em{\sc i\kern-.025em b}\kern-.08em
    T\kern-.1667em\lower.7ex\hbox{E}\kern-.125emX}}

\begin{document}

\title{Large Language Models in Thematic Analysis: Prompt Engineering, Evaluation, and Guidelines for Qualitative Software Engineering Research}


\author{
Cristina Martinez Montes$^1$, Robert Feldt$^1$, Cristina Miguel Martos$^2$, Sofia Ouhbi$^3$, Shweta Premanandan $^3$, Daniel Graziotin$^4$, \\
$^1$\textit{Chalmers University of Technology and University of Gothenburg}, $^2$\textit{Universidad Ramon Llull}
$^3$\textit{Uppsala University}
$^4$\textit{University of Hohenheim}\\
montesc@chalmers.se, robert.feldt@chalmers.se, cristina.miguel@iqs.url.edu, sofia.ouhbi@it.uu.se, shweta.premanandan@it.uu.se, graziotin@uni-hohenheim.de}


\maketitle

\begin{abstract}
As artificial intelligence advances, large language models (LLMs) are entering qualitative research workflows--yet no reproducible methods exist for integrating them into established approaches like thematic analysis (TA), one of the most common qualitative methods in software engineering research. Moreover, existing studies lack systematic evaluation of LLM-generated qualitative outputs against established quality criteria.

We designed and iteratively refined prompts for Phases 2–-5 of Braun and Clarke's reflexive TA, then tested outputs from multiple LLMs against codes and themes produced by experienced researchers. Using 15 interviews on software engineers' well-being, we conducted blind evaluations with four expert evaluators who applied rubrics derived directly from Braun and Clarke's quality criteria. Evaluators preferred LLM-generated codes 61\% of the time, finding them analytically useful for answering the research question. However, evaluators also identified limitations: LLMs fragmented data unnecessarily, missed latent interpretations, and sometimes produced themes with unclear boundaries.

Our contributions are threefold. First, A reproducible approach integrating refined, documented prompts with an evaluation framework to operationalise Braun and Clarke’s reflexive TA. Second, an empirical comparison of LLM- and human-generated codes and themes in software engineering data. Third, guidelines for integrating LLMs into qualitative analysis while preserving methodological rigour ---clarifying when and how LLMs can assist effectively and when human interpretation remains essential.

\end{abstract}


\section{Introduction} \label{sec:intro}

Recent advances in artificial intelligence (AI) have enabled Large Language Models (LLMs) to process vast amounts of text and uncover complex patterns with remarkable speed \cite{roberts2024artificial, tai2024examination, bano2023ai}. These capabilities make LLMs especially promising for qualitative data analysis (QDA).

While software engineering (SE) research has traditionally emphasized quantitative and experimental methods \cite{basili2007empirical, kitchenham2002preliminary, perry2000empirical}, the discipline is increasingly recognized as social, multidisciplinary, and deeply human \cite{harrison1999directions, lenberg2015behavioral, stol2018abc}. Understanding software development demands attention to real-world contexts \cite{meyer2013empirical} and to the interplay of technical, managerial, and organizational factors \cite{dybaa2011qualitative}. As a result, qualitative methods---such as grounded theory, thematic analysis (TA), and content analysis---have gained growing traction in SE research \cite{stol2018abc,lenberg2024qualitative}.


As qualitative methods gain prominence and LLMs emerge as tools for qualitative data analysis (QDA), research in this area has expanded rapidly \cite{de2023performing, bano2024large, wen2025leveraging}. Yet, several limitations and concerns persist.

\textbf{Limited transparency and explainability.} Many SE studies fail to report key details such as prompts, model configurations, or evaluation procedures, hindering transparency, reproducibility, and interpretability \cite{bennis2025advancing}.

\textbf{Lack of systematic evaluations on SE data.} Most existing work focuses on comparing LLM outputs with human-coded results \cite{bano2023ai}, overlooking other vital quality dimensions---such as the coherence and depth of themes, the transparency of coding decisions, and the usefulness of the resulting insights for addressing research questions.

\textbf{Insufficient methodological rigour.} Many studies adopt ad hoc or poorly justified procedures, applying LLMs to QDA without grounding their approach in established methodological frameworks. Few validate their processes against accepted qualitative standards or employ systematic checks for reliability and interpretive depth \cite{byrne2022worked, de2024performing}. In contrast, our study followed Braun and Clarke’s reflexive thematic analysis (TA) framework \cite{braun2021thematic} and provided detailed documentation of each step.

\textbf{Narrow model scope.} Prior studies often rely on a single LLM, which limits the generalisability of their findings. This also leaves open questions about how model choice influences coding quality and interpretive outcomes. Our study addressed this by evaluating several leading models. We tested different LLMs and evaluated their performance supporting more robust conclusions about LLM performance.

Beyond methodological shortcomings, researchers have also highlighted ethical and privacy risks, particularly when working with sensitive data \cite{bano2024large}.

This study addresses these gaps by empirically evaluating the use of LLMs (ChatGPT 03 mini, GPT-4o, Gemini 2.5 Pro, and Claude 4 Sonnet) in TA. TA is one of the most widely applied methods for qualitative research in SE \cite{lenberg2024qualitative}. We systematically compared human and LLM-generated codes and themes between March and July 2025. We evaluated \footnote{The evaluators' role was limited to only reviewing and rating the codes and themes. All prompting, model runs, and refinement of outputs were carried out by the first and second authors.} them using rubrics derived from Braun and Clarke's reflexive thematic analysis framework \cite{braun2021thematic}. To enhance reliability, we iteratively refined and fully disclosed prompts to ensure transparency and reproducibility. 

Our study also advances methodological practice by operationalising Braun and Clarke's reflexive TA for use with LLMs. We offer practical guidelines for integrating AI in ways that support rather than replace researcher reflexivity. 

This study answered the following research question:\\

\textbf{RQ: To what extent can LLMs perform phases of reflexive TA in a way that aligns with established qualitative research standards?}\\

Our main contributions are:
\begin{itemize}
    \item A reproducible framework combining prompt design and rubric-based evaluation for applying reflexive thematic analysis with LLMs.
    \item An empirical comparison of LLM- and human-generated codes and themes in software engineering data.
    \item Guidelines for integrating LLMs into qualitative analysis workflows to enhance efficiency while preserving reflexivity and methodological rigour.
\end{itemize}

This study adds to the methodological discussions in empirical SE by clarifying the possibilities and the limits of AI-assisted qualitative research.

\section{Background and Related Work} \label{sec:related}
This section provides the background of the study on Thematic Analysis and the related work on using AI and LLMs in qualitative data analysis, particularly in TA.

\subsection{Qualitative Data Analysis in Software Engineering}

QDA in SE allows a deep exploration of non-technical aspects of software development \cite{seaman1999qualitative}. Researchers find patterns, meanings, and insights by systematically interpreting rich, non-numerical data \cite{treude2024qualitative, seaman1999qualitative}. QDA is particularly useful for gaining a deep understanding of software processes, tool use, and organisational or technical settings. In these cases, interpretation and contextual insights are crucial for advancing theory and practice in SE. Additionally, SE qualitative datasets often blend technical artefacts. For example, code review comments, architecture decision records, incident chats with human-centred sources (interviews, field notes). This emphasises the importance of scale management, cross-analyst consistency, and traceable decision trails for credibility.

\subsubsection{Thematic analysis}\label{sec:TA}

We chose TA for this study because it is one of the most popular data analysis methods within SE research \cite{lenberg2024qualitative, dittrich2007editorial}. We adapted the version by Braun and Clarke, Reflexive Thematic Analysis \cite{braun2021thematic}, for collaboration with LLMs. The six phases are detailed next:

\begin{itemize}
    \item Phase 1. Familiarisation with the Data: Reading and re-reading the data is required to understand the content fully.
    \item Phase 2. Generating Initial Codes: The goal is to systematically identify and label data segments that present ideas or concepts that could help answer the research question.
    \item Phase 3. Generating Initial Themes:  The aim is to cluster codes with similar meaning into candidate themes with patterns and broad ideas.
    \item Phase 4. Developing and Reviewing Themes: It extends phase 3 and does a vital check to review and explore the initial clusterings to find a better pattern development based on the research question.
    \item Phase 5. Refining, Defining and Naming Themes: The final themes are refined by determining the structure and flow of the analysis. It also requires writing a definition and naming them in a way that represents their content and central idea.
    \item Phase 6. Writing up the analysis: Involves explaining the findings in themes to answer the research questions coherently and effectively. It also includes selecting and using extracts from the data to illustrate the core parts of the themes.
\end{itemize}

\subsection{Early AI and ML in Qualitative Data Analysis}

Recent advancements in AI have sparked a growing interest in leveraging its application to qualitative data analysis. For example, Liew et al.\cite{liew2014optimizing} proposed a method that involves natural language processing (NLP) and machine learning (ML) to generate initial codes, which are subsequently refined by human input. Similarly, various other studies have used NLP to derive potential codes \cite{crowston2012using, crowston2010machine, grimmer2013text, lewis2013content}. Other studies have instead outlined challenges of implementing ML for qualitative coding \cite{chen2018using}.

In a similar line, Towler et al. \cite{towler2023applying} proposed Machine-Assisted Topic Analysis (MATA). This is an NLP approach that combines human input with automated analysis to summarise text patterns more efficiently. MATA's features make it valuable for qualitative researchers handling large datasets. Compared to traditional TA, MATA is less time- and resource-intensive, aiding in early familiarisation and coding. A similar tool is LaMa \cite{bogachenkova2023lama}, short for machine labelling. It is a web application that facilitates the handling and tracking of labels and changes. It makes it easy for researchers to group labels, create themes, and collaborate. However, unlike our study, these tools do not generate codes or themes. Our approach allows the LLM to propose codes and themes. Having initial codes provides an initial analytical layer for researchers and supports a more comprehensive initial capture of the data. 

While traditional supervised and unsupervised ML techniques have been widely employed in qualitative analysis \cite{renz2018two, nelson2020computational, gauthier2022computational, qiao2025thematic, sinha2024role, yan2024human, nguyen2025chatgpt}, significant gaps remain in addressing challenges such as privacy, model bias, quality control, and reproducibility \cite{bano2024large}. In this study, we addressed these concerns by implementing stepwise human oversight, transparent prompt design, and systematic evaluation procedures.

\subsection{LLMs in Qualitative Analysis}

Several studies have investigated LLMs' role in supporting QDA. A first set of works has explored the potential of LLMs for inductive or deductive coding when doing TA.
De Paoli \cite{de2023performing} explored the application of ChatGPT 3.5-Turbo to conduct inductive TA in semi-structured interviews. Using open-access interviews previously analysed by human researchers, De Paoli showed the capacity of LLMs to infer main themes from prior research contexts. Moreover, the study emphasises the LLMs' capability to identify relevant themes that might have eluded human analysts. However, methodological rigour across TA phases was not assessed. Our study evaluated analytic quality and coherence across Phases 2–5 using a rubric-based assessment and disclosed prompts.
In the same way, Xiao et al.\cite{xiao2023supporting} focused on evaluating the deductive coding agreement between LLMs and human analysts. They also investigated how the design of prompts influences analysis outcomes. Their focus was on agreement and prompt effects; we additionally test interpretive depth, theme coherence, and reflexive alignment with the RQs.

Wen et al. \cite{wen2025leveraging} extended this line of inquiry, testing LLMs to perform inductive and deductive coding with a large-scale case study in the charity sector. They achieved strong semantic alignment with human coding and sentiment analysis, yet also found inconsistencies in excerpt extraction and the heavy need for human validation. In our approach, We mitigated these issues by coding interviews segment by segment and integrating continuous human feedback rather than relying on post-hoc validation.

Beyond coding, researchers have also explored how LLMs might contribute to higher-level analytic tasks. Tabone and de Winter \cite{tabone2023using} showed that GPT-generated sentiment ratings and summaries were often consistent with human outputs. However, results varied depending on the prompt, and GPT sometimes produced themes absent in human analyses. Their outputs may be useful, but they risk inconsistency or distortion without systematic evaluation. Our work focuses on the analytical core of TA. We incorporated human feedback to ensure depth and trustworthiness while maintaining the final interpretative step as fully human as possible.

These previous studies have proven that LLMs can assist with coding, summarisation, and collaborative workflows. Still, most studies focused on isolated tasks, expressed prompt sensitivity, or stopped short of assessing analytic rigour. Our study contributes to the existing body of literature and addresses these limitations. We applied human and rubric-based evaluations between steps and embedded transparency and reflexivity in the prompts.

 \begin{figure*}[!h!]
    \centering
    \includegraphics[width=1\linewidth]{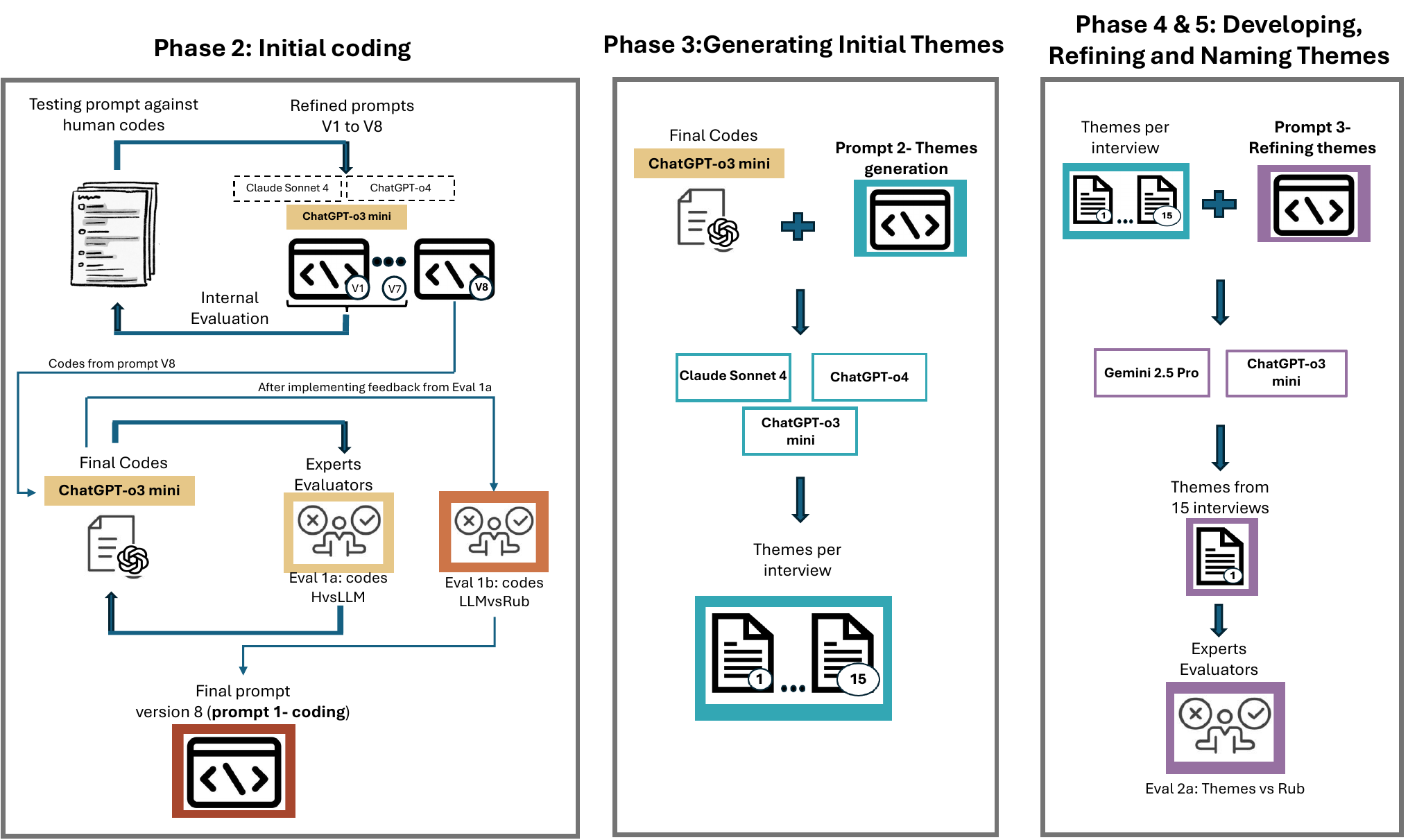}
    \caption{Study's methodology overview. The dataset was coded in Phase 2 with Prompt 1, evaluated by experts, and refined iteratively to obtain the final version (v8). In Phase 3, Prompt 2 generated themes per interview, and in Phases 4–5, Prompt 3 created themes across 15 interviews. Final outputs from different model pipelines were again reviewed by experts using Braun and Clarke's quality criteria.}
    \label{fig:Method}
\end{figure*}

\subsection{Generative AI Tools and Frameworks in QDA}

Researchers have started to propose tools and frameworks to integrate LLMs as collaborators in QDA.  For example, CollabCoder \cite{gao2024collabcoder} is a one-stop, end-to-end workflow used for inductive coding. It offers AI-generated code suggestions, facilitates iterative discussions using quantitative metrics, and provides primary code suggestions for creating codebooks. However, this tool focuses only on one specific type of coding (inductive) and does not consider the RQs, which are the drivers in qualitative analysis. In our study, our prompt included the RQ. Moreover, the tool only addresses three analysis steps; in contrast, our study also covers the themes' creation step. In addition, CollabCoder run under certain assumptions that are not always true when conducting TA (the presence of two coders, raw data consisting of semantically distinct units,  and data segments each conveying a single meaning). Our tool is not tied to a certain number of researchers and can handle any type of raw data. For example, interviews, where a single segment can have more than one meaning.

In a similar line, Gebreegziabher et al. proposed Patat \cite{gebreegziabher2023patat}. This tool learns patterns from user-annotated codes and recommends new ones. It also creates codebooks and helps the user learn data characteristics. Patat improves explainability compared to CollabCoder by showing users what the model has learned. In line with this, our prompt also explains codes and themes. A limitation of Patat is that it supports only one user at a time, which is problematic since qualitative analysis is typically conducted collaboratively by multiple researchers. Although it offers several features, its creators acknowledge that this makes the tool complex and challenging to learn. Unlike Patat, our tool supports collaborative analysis, extends across multiple TA phases, and evaluates analytic quality using reflexive rubrics.

 Drápal and Savelka's \cite{drapal2023using} framework, designed for legal experts to collaborate with OpenAI's GPT-4 model, is the closest to our study, although in a different field. They covered phases 2 (generating initial codes), integrating users' feedback, 3 (searching for themes), and 4 (generating initial themes). However, their work was restricted to a legal domain and stopped short of evaluating methodological. By contrast, our study not only covered the same steps but also added explanations of the model decisions to improve trust. We also tested several models in parallel for comparison, and incorporated extra tasks such as defining themes, mapping codes to themes, and flagging sensitive segments. Furthermore, unlike Drápal and Savelka's framework, we embedded step-by-step evaluations with human experts based on Braun and Clarke's guidelines. Additionally, we refined prompts iteratively throughout the process to ensure transparency and reproducibility.

 Taken together in our study, we integrate systemic human evaluation using tailored rubrics based on Braun and Clarke's TA framework, making the assessment rigorous. Furthermore, we engineered prompts for steps 2 to 5 in reflexive TA and fully disclosed them, thereby advancing transparency and methodological clarity beyond prior work.

\section{Methodology} \label{sec:methodology}

This section elaborates on the steps we followed in the study, the design, implementation, and evaluation. We adopted a mixed-methods approach, combining qualitative and quantitative data collection. We aimed to assess the capabilities of LLMs in conducting TA following Braun and Clarke's framework.

\begin{table*}[h]
\centering
\caption{Rubric used to evaluate the quality of initial codes generated by LLMs during Phase 2 of thematic analysis. The rubric includes eight criteria adapted from Braun and Clarke's guidelines, each rated on a 4-point scale from ``Poor" (1) to ``Excellent" (4).}
\begin{tabular}{p{3cm}p{3.25cm}p{3.25cm}p{3.25cm}p{3.25cm}}
\hline
\textbf{Criteria} & \textbf{Excellent (4)} & \textbf{Good (3)} & \textbf{Fair (2)} & \textbf{Poor (1)} \\ 
\hline

\textbf{Clarity of Meaning} & Codes are exceptionally clear, specific, and unambiguous, capturing distinct and well-defined meanings within the data. & Codes are mostly clear and specific, with minor ambiguities that do not hinder understanding. & Codes are somewhat unclear, leading to some ambiguity in meaning and interpretation. & Codes are unclear, vague, or ambiguous, failing to capture distinct meanings. \\ 
\hline

\textbf{Relevance to Research Question} & Codes are highly relevant, directly addressing and reflecting the research question with strong alignment to the data. & Codes are mostly relevant, with a few minor deviations from the research question. & Codes are somewhat relevant but fail to capture important aspects of the research question fully. & Codes are largely irrelevant, showing little to no connection to the research question. \\ 
\hline

\textbf{Balance of Latent and Semantic Meanings} & Codes effectively capture surface-level and deeper meanings, demonstrating a strong balance between the two. & Codes capture either surface-level or deeper meanings effectively, but not both equally. & Codes focus primarily on surface-level meanings, neglecting deeper insights. & Codes fail to capture both surface-level and deeper meanings, lacking depth. \\ 
\hline

\textbf{Specificity} & Codes are precise, capturing narrow and distinct meanings that do not overlap with other codes. & Codes are mostly precise, with occasional overlaps that may cause some confusion. & Codes lack precision, with significant overlaps leading to unclear distinctions. & Codes are imprecise and broad, with substantial overlap, making distinct meanings unclear. \\ 
\hline

\textbf{Potential for Theme Development} & Codes provide a robust foundation for meaningful theme development, reflecting diverse insights. & Codes mostly support theme development, but may lack some diversity in insights. & Codes provide limited potential for theme development, lacking diversity and clarity. & Codes do not support theme development, reflecting a narrow range of insights. \\ 
\hline

\textbf{Alignment with Data} & Codes are closely aligned with the dataset content, accurately reflecting the meaning of the data. & Codes are mostly aligned, with minor discrepancies in reflecting the data’s meaning. & Codes show some misalignment with the data, leading to inaccurate representations. & Codes are poorly aligned, failing to reflect the dataset’s meaning accurately. \\ 
\hline

\textbf{Good Labels} & Code labels offer a concise, pithy, and insightful shorthand for broader ideas, enhancing understanding. & Code labels are mostly concise and insightful, but some could be improved for clarity. & Code labels are somewhat vague or overly broad, lacking clarity in labelling. & Code labels are unclear and lengthy, failing to provide effective shorthand for ideas. \\ 
\hline

\textbf{Explanation of Interview Segment Selection} & The explanation is exceptionally clear and logical. It directly and explicitly connects the coded interview segment to the research questions and convincingly demonstrates its importance to the main topic. The reasoning is thorough and well-justified, leaving no ambiguity. & The explanation is clear and mostly logical, with minor areas that could be more detailed. It connects the coded interview segment to the research questions and shows its importance to the main topic; however, the link could be more explicit or compelling. & The explanation is somewhat clear but confusing in parts. It partially connects the coded interview segment to the research questions and mentions its importance to the main topic; however, the relevance and justification are weak or underdeveloped. & The explanation is unclear, disjointed, or difficult to follow. It fails to connect the coded piece to the research questions or demonstrate its importance to the main topic. The reasoning is missing, vague, or irrelevant.
\\ 
\hline
\end{tabular}
\label{tab:RubcodesEval}
\end{table*}

\subsection{Dataset}

Our dataset consisted of 15 semi-structured interviews that explored factors that influence the well-being of software engineers, collected from a previous study \cite{montes2024factors}. Each interview lasted between 40 and 75 minutes and was audio-recorded, transcribed, and anonymised. The transcription resulted in 177 pages in Word using a font size of 12.
Two researchers previously inductively coded the interviews following Braun and Clarke's guidelines \cite{braun2021thematic}. We used this set of codes (human-generated) to compare with the LLM codes at multiple stages. 

\subsection{Study Design: Mapping TA Phases to Human vs LLM Roles}

The first two authors (referred to as ``we” from now on) structured the study to match Phases 2 to 5 of Braun and Clarke's Reflexive TA framework (see Figure \ref{fig:Method} for the overview of the experiment):

\begin{itemize}
    \item Phase 2: Generating Initial Codes
    \item Phase 3: Generating Initial Themes (per interview)
    \item Phase 4: Developing and Reviewing Themes
    \item Phase 5: Refining, Defining, and Naming Themes
\end{itemize}

\begin{figure}
    \centering
    \includegraphics[width=1\linewidth]{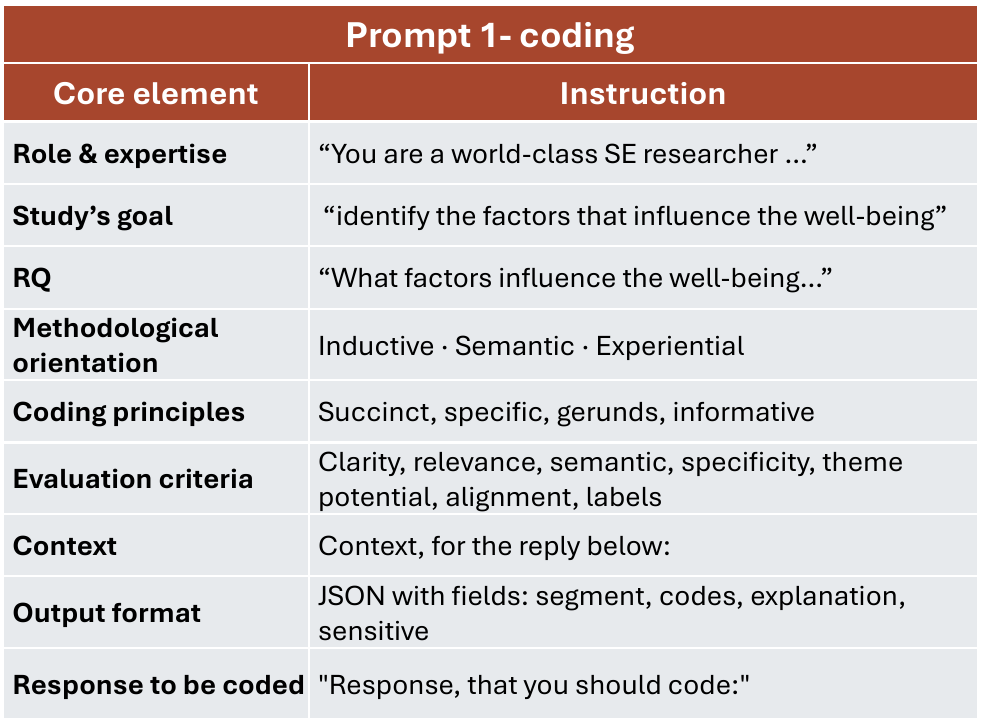}
    \caption{Core elements and short examples of prompt 1, final version for creating codes.}
    \label{fig:prompt1}
\end{figure}

\begin{figure}
    \centering
    \includegraphics[width=1\linewidth]{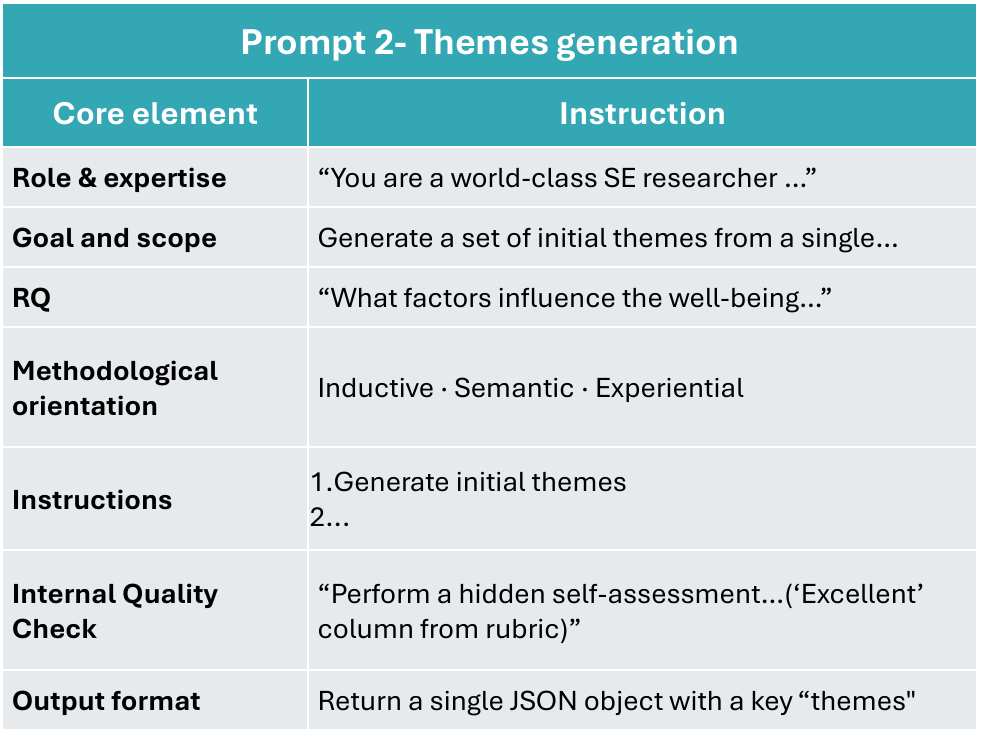}
    \caption{Core elements and short examples of prompt 2 for creating themes per interview.}
    \label{fig:prompt2}
\end{figure}

\begin{figure}
    \centering
    \includegraphics[width=1\linewidth]{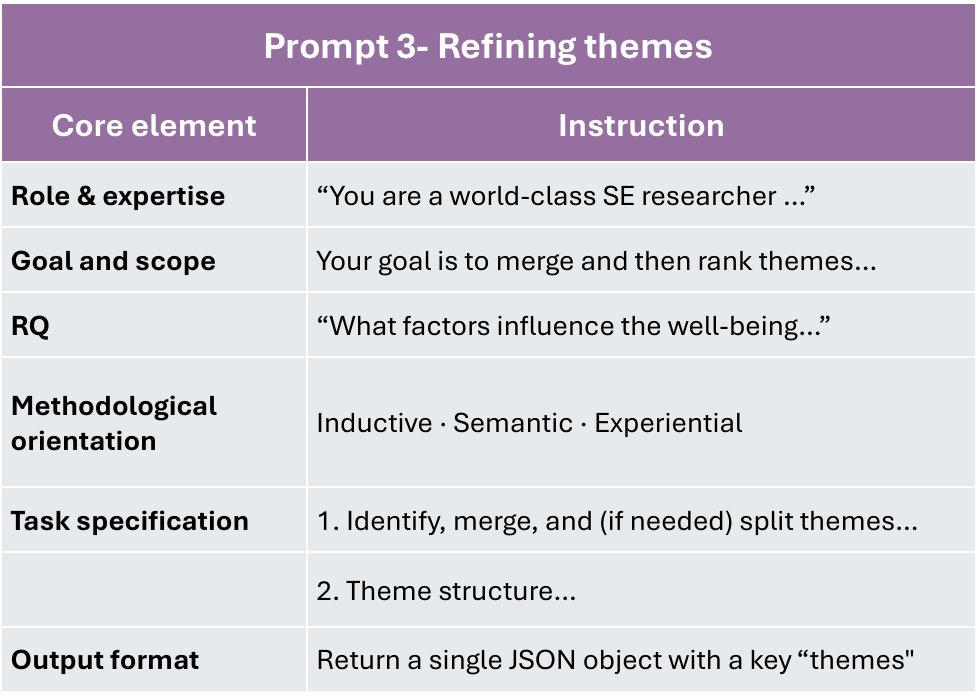}
    \caption{Core elements and short examples of prompt 3 for creating themes, including 15 interviews.}
    \label{fig:prompt3}
\end{figure}
 
We left phases 1 (Familiarisation)  and 6 (Producing the Report) entirely human-led. However, we implicitly embedded familiarisation within the initial prompting of Phase 2 by providing the LLM interview and problem context. Phase 6 requires human interpretation, as it involves contextualising themes and ensuring alignment with the RQs and qualitative standards. The final write-up must be grounded in a solid theoretical foundation and developed through a rich, rigorous interpretative process \cite{braun2021thematic}. This makes the task more challenging for LLMs. We used four LLMs across the study. Table \ref{tab:llm_overview} presents the characteristics of each model and indicates the phases in which they were implemented.
 
\begin{table*}[h]
\centering
\caption{Rubric to evaluate the quality of themes. The rubric includes eight criteria adapted from Braun and Clarke's framework: coherence, relevance, boundary clarity, data support, definition, naming, analytical contribution, and use of subthemes. Each criterion is rated on a 4-point scale from ``Poor" (1) to ``Excellent" (4).}
\begin{tabular}{p{3cm}p{3.25cm}p{3.25cm}p{3.25cm}p{3.25cm}}
\hline
\textbf{Criteria} & \textbf{Excellent (4)} & \textbf{Good (3)} & \textbf{Fair (2)} & \textbf{Poor (1)} \\ 
\hline

\textbf{Central Organising Concept and Conceptual Coherence} & The theme has a coherent, clear, distinct, and well-defined central organising concept that seamlessly ties all data and codes. & The theme has a central organising concept that ties most data and codes together, with minor gaps in coherence. & The theme has a central organising concept, but is somewhat vague or inconsistently applied. & The theme lacks a coherent, clear central organising concept.  
\\ 
\hline

\textbf{Meaningfulness and Relevance} & The theme captures something highly meaningful and relevant to the research questions. &  The theme captures something meaningful and relevant, but the connection to the research questions could be more explicit or detailed.	& The theme captures some meaningful aspects, but its relevance to the research questions is unclear or weakly argued. &	The theme does not capture anything meaningful or relevant to the research questions. 
\\ 
\hline

\textbf{Clarity of Boundaries} & The theme has clear and well-defined boundaries. It is distinct from other themes, with no overlap or confusion. &	The theme has mostly clear boundaries, with minor overlaps or ambiguities that do not significantly detract from its distinctiveness.	& The theme has somewhat unclear boundaries, with noticeable overlaps or ambiguities that weaken its distinctiveness. &	The theme lacks clear boundaries. It overlaps significantly with other themes or is too broad/vague to be distinct.

\\ 
\hline

\textbf{Data Support and Evidence} & Strongly supported by meaningful and sufficient data, with diverse yet coherent evidence.	& Supported by sufficient data, but some data points could be more strongly aligned.	& Partially supported by data, but with gaps or inconsistencies in alignment.	& Lacks sufficient or meaningful data support; data are sparse, irrelevant, or misaligned

\\ 
\hline

\textbf{Theme Definition} & The definition clearly outlines the theme's central organising concept, boundaries, and uniqueness.	& The definition outlines the central concept and boundaries, but could be sharper. &	The definition partially explains the central concept and boundaries but lacks depth or clarity. &	The definition is missing, unclear, or fails to explain the theme's central concept, boundaries, or uniqueness.
\\ 
\hline

\textbf{Theme Name} & The name is informative, concise, and catchy. &	The theme name is clear and informative, but could be more concise or engaging. &	The theme name is somewhat unclear or generic. &	The theme name is vague or uninformative. 
\\ 
\hline

\textbf{Contribution to Overall Analysis} &  The theme significantly and uniquely contributes to the overall analysis. It adds depth, insight, and clarity to the research questions and findings.	& The theme contributes to the overall analysis, but its unique contribution could be more explicitly stated or developed.	& The theme contributes partially to the overall analysis, but its role is unclear or underdeveloped.	& The theme does not contribute to the overall analysis. It seems redundant, irrelevant, or disconnected from the research questions and findings.

\\ 
\hline

\textbf{Subthemes (if existent)} & Subthemes are conceptually clear, non-overlapping, and each captures a distinct facet of the central organising concept. They enhance the narrative's meaning. &	Subthemes are relevant and mostly well-aligned with the central concept. Minor overlap or lack of distinctness, but they still support theme clarity. &	Subthemes are weakly connected to the central theme or to each other. They show some redundancy or confusion, weakening coherence. &	Subthemes are misaligned, redundant, vague, or unnecessary. They add little value and may introduce confusion.
\\ 
\hline

\end{tabular}
\label{tab:RubricThemes}
\end{table*}

\subsection{Prompting Strategy, Application and Evaluation}
We constructed, tested, evaluated, refined and rewrote several prompts for each phase (see the prompts in the online appendix \cite{martinez_montes_2025_17401526}). The evaluations combined quantitative tools, such as rubrics, with qualitative data from the evaluators' comments. 

We created two rubrics based on Braun and Clarke's quality benchmarks for TA outputs. The rubric for evaluating the quality of the codes from Phase 2 is presented in Table \ref{tab:RubcodesEval}. It follows Braun and Clarke's guidelines for Phase 2 of reflexive thematic analysis (`generating initial codes') \cite{braun2021thematic}. It includes eight criteria: clarity of meaning, relevance to the research question, balance of latent and semantic meanings, specificity, potential for theme development, alignment with data, quality of code labels, and explanation of interview segment selection. The rating has a four-point scale from ``Poor" (1) to ``Excellent" (4). 

Meanwhile, the rubric for evaluating the quality of themes is shown in Table \ref{tab:RubricThemes}. It defines eight criteria: coherence, relevance, boundary clarity, data support, definition, naming, analytical contribution, and use of subthemes. Each criterion is rated on a four-point scale from ``Poor" (1) to ``Excellent" (4). 

The rubrics facilitated a systematic, transparent, and comparable evaluation of codes and themes against qualitative research standards. The combination of data supported that assessments were standardised. The whole process we followed is explained next:\\

\subsubsection{\textbf{Prompting Strategy}}

Creating and refining prompts was an iterative process that involved testing, assessing, implementing feedback, and retesting. 

The final prompts were intentionally lengthy, as we included clear and consistent guidance throughout the analytic steps. However, the trade-off is that long prompts might reduce flexibility, increase computational cost, and make replication or adaptation more difficult. We included the full prompts in the online appendix \cite{martinez_montes_2025_17401526}. In the paper, only the core parts are presented for brevity and overview. We elaborate in each prompt next:

\begin{table}[t]
\centering
\caption{LLMs characteristics and dates of access per TA phase. The Phase is represented by ``P"+ number.}
\setlength{\tabcolsep}{3pt}
\renewcommand{\arraystretch}{1.15}
\begin{tabular}{@{}%
  >{\raggedright\arraybackslash}p{0.24\columnwidth}%
  >{\raggedright\arraybackslash}p{0.30\columnwidth}%
  >{\raggedright\arraybackslash}p{0.40\columnwidth}@{}}
\toprule
Model & Settings (Temp, Max Tokens) & Phases and Dates of Access \\
\midrule
ChatGPT o3-mini &
  Temp = default,\newline Max tokens = 200k &
  P:2, date 21 of May, 25\newline
  P:3, date: 6 of July, 25\newline
  P:4\&5, date 9 of july, 25 \\
GPT-4o &
  Temp = default,\newline Max tokens = 64,000 &
  P:2, date 21 of March, 25\newline
  P:3, date: 6 of July, 25 \\
Claude Sonnet 4 &
  Temp = default,\newline Max tokens = 64,000 &
  P:2, date 21 March, 25\newline
  P:3, date: 23 of May, 25 \\
Gemini 2.5 Pro &
  Temp = default,\newline Max tokens = 1,048,576 &
  P:3, date: 8 of July, 25\newline
  P:4\&5, date date 9 of july, 25 \\
\bottomrule
\end{tabular}
\label{tab:llm_overview}
\end{table}

\paragraph{Prompt 1- coding }~went through eight versions. We began with short instructions for the LLM on generating codes from interview responses in line with Braun Clarke's reflexive TA framework. Preliminary outputs had issues, including overly broad codes, hallucinated segments, and insufficiently descriptive labels. We addressed these by adding key elements and conducting internal dry runs with ChatGPT o3-mini, Claude Sonnet 4, and GPT-4o. The final prompt, version 8, generated the codes to be sent to the evaluators. This version requested the model to act as a qualitative researcher, identifying meaningful data segments in participants' responses and code labels. Each coded segment included the verbatim quote, a brief explanation of its relevance, and a note on any sensitive information. The prompt requested the LLM to write the specific interview number and line from which the code was taken. With this, we made sure that all codes were real and avoided hallucinations. We performed quality checks following this step by cross-verifying the excerpts with the original transcripts to confirm accuracy and consistency. The prompt also included a coding quality check using the ``Excellent" column from the rubric in Table \ref{tab:quali_rubric_codes}. Additionally, it specified detailed methodological guidance and structured JSON output for later theme development. 
Since the prompt was long, 1485 tokens in total, we show an overview in Figure \ref{fig:prompt1}. See the online appendix \cite{martinez_montes_2025_17401526} for the full prompt.

\paragraph{Prompt 2- Themes generation}~instructed the LLM to act as a qualitative software engineering researcher applying Braun and Clarke's TA. It requested the production of initial themes from the coded interview by Prompt 1-coding. The model received coded segments and must derive themes that directly address the study's RQ. It requested to group related codes into coherent, data-driven themes and sub-themes. It also asked to provide concise, meaningful theme names and write detailed definitions. The prompt included methodological constraints, a style example for depth and tone, and an internal self-check rubric based on the ``Excellent" column from the rubric in Table \ref{tab:RubricThemes}. All of this presented in a JSON output structure. 

Prompt 2 tested whether the LLM could generate themes across a full transcript. This step ensured that the model could handle larger amounts of qualitative data while maintaining alignment with Braun and Clarke's TA framework. We tested the prompt with ChatGPT o3-mini, GPT-4o and Claude Sonnet 4. The goal was to assess each model's ability to identify patterns across individual interviews before proceeding to the entire set. The prompt consists of 1280 tokens. The overview is shown in Figure~\ref{fig:prompt2}, and the full prompt is provided in the online appendix \cite{martinez_montes_2025_17401526}.

\paragraph{Prompt 3- Refining themes}~was to generate themes for all 15 interviews, building on the sets of themes produced by Prompt 2. Whereas Prompt 2 generated themes per interview, Prompt 3 synthesised patterns across the full dataset. The aim was to produce a coherent set of candidate themes that answer the RQ. 

This prompt asked the LLM to do the theme aggregation and refinement stage of TA. It used the initial themes from individual interviews and guided the model to merge themes. It also requested to produce a coherent, ranked set of overarching themes. The model assessed the significance of each theme based on its explanatory power, frequency, and diversity of supporting evidence. It also assigned ranks to high/medium/lower tiers, and records the source themes. For each theme and sub-theme, a detailed, human-quality definition was required. It described its central organising concept, boundaries, uniqueness, and contribution to answer the RQ.

We used the prompt in different LLM pipelines to produce 5 sets of candidate themes. Prompt~3 has 1319 tokens. See Figure \ref{fig:prompt3} for an overview and the online appendix \cite{martinez_montes_2025_17401526} for the full text). \\

\subsubsection{\textbf{Initial Coding (Phase 2)}} 
We used the final version (V8) of prompt 1 to generate the codes for TA phase 2. The codes were then assessed and evaluated.  

\begin{figure}
    \centering
    \includegraphics[width=1\linewidth]{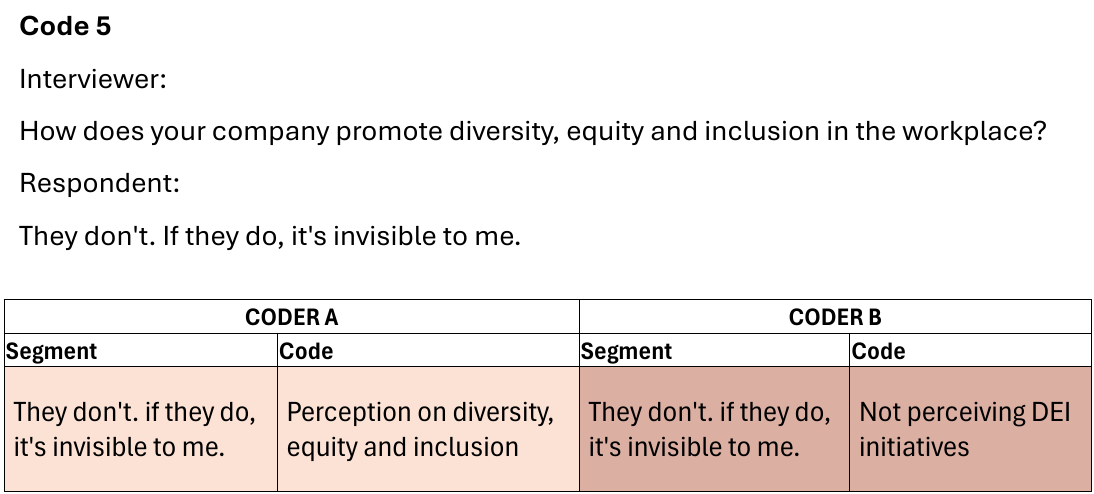}
    \caption{Example of human- and LLM-generated codes presented in the blind evaluation. Evaluators chose between coder A and coder B and provided a brief justification for their decision. Here, the whole reply was coded, an uncommon case, shown for brevity and clarity.}
    \label{fig:eval_code_example}
\end{figure}

\textbf{Evaluation.} We evaluated this phase in two steps:

\paragraph{Eval 1a: codes HvsLLM:\label{1a}}~First, we compared human vs LLM. A total of 96 sets of codes were presented to evaluators randomly and blindly. Each set had between 1 and 4 codes. The sets were formed by human and LLMs codes from different interview segments. Evaluators received the interview transcript and the RQ to gain a comprehensive understanding of the context. They also received an online survey, with each question addressing one set. Evaluators could see both sets (human and LLM) simultaneously. Figure \ref{fig:eval_code_example} shows how evaluators saw each set and answered the survey. Then they assessed each set, selecting which codes better captured meaning and addressed the RQ. They also provided written justifications for their choices. Each set consisted of the interview question, the participant's response, and the corresponding code (human and LLM-generated).

\paragraph{Eval 1b: codes LLMvsRub:}~Second, we compared LLM vs rubric. After refining the prompt and generating additional codes, 24 sets of LLM-only codes were evaluated using a rubric (see Table \ref{tab:RubcodesEval}). The rubric criteria reflect Braun and Clarke's qualities as essential for high-quality code generation. The four evaluators independently rated the codes and provided written feedback, yielding quantitative scores and qualitative insights.\\

\subsubsection{\textbf{Generating, Reviewing, Refining and Naming Themes (Phases 3-5)}} \label{sec:phase2:b}

For phase 3 (generation of initial themes), we used prompt 2 to generate the first themes per interview in ChatGPT o3-mini, GPT-4o, Claude 4 Sonnet models. This phase was not evaluated because it generated per-interview themes. These were intermediate analytic artefacts that required further refinement in Phases 4 and 5 before they could be meaningfully interpreted. Evaluating Phase 3 alone would not provide valid insight into analytical quality, as reflexive thematic analysis treats theme development as an iterative and cumulative process. Consequently, we focused our evaluation efforts on the more stable outputs from Phases 2 and 4 \& 5, which better reflect interpretive rigour and final analytic quality. 

Then, we used prompt 3 to cover phases 4 and 5 (refined, defined and named the themes). We used five different LLM pipelines (see Methodology, Figure \ref{fig:Method}) across all 15 interviews. The pipelines combined ChatGPT o3-mini, GPT-4o, Claude 4 Sonnet, and Gemini 2.5 Pro.

\begin{table}[t]
\centering
\caption{Pipelines tested in phases 4 and 5 to create the final themes. Pipeline 5 produced the best set of themes to answer the RQ}
\label{tab:pipelines}
\begin{tabular}{lccc}
\toprule
\textbf{Pipeline} & \textbf{Phase 2} & \textbf{Phase 3} & \textbf{Phase 4 \& 5} \\
\midrule
P1 & ChatGPT o3-mini & Gemini 2.5 Pro & Gemini 2.5 Pro \\
P2 & ChatGPT o3-mini & ChatGPT o3-mini & Gemini 2.5 Pro \\
P3 & ChatGPT o3-mini & ChatGPT o3-mini & ChatGPT o3-mini \\
P4 & ChatGPT o3-mini & Claude Sonnet 4  & ChatGPT o3-mini \\
P5 & ChatGPT o3-mini & Claude Sonnet 4  & Gemini 2.5 Pro \\
\bottomrule
\end{tabular}
\end{table}

Five sets of themes were generated and assessed for completeness, code quality, structural coherence, and the presence of subthemes. Pipelines that omitted interviews or lacked a clear structure were excluded. The best-performing pipeline presented all the assessment criteria. Table \ref{tab:pipelines} shows all the pipelines we used to generate theme sets. Then we selected the most promising set, produced by P5, to be scored by the evaluators. Nine distinct themes formed the resulting set of themes.

\textbf{Eval 2a: Themes vs Rub:} The final set of themes was independently evaluated by three experts. They used the rubric in Table \ref{tab:RubricThemes} and provided free-text comments for overall feedback.

\subsection{Data analysis}

For the quantitative results from the rubrics, we summed the evaluators' scores for each criterion and calculated averages. We conducted a content analysis for the qualitative feedback (free-text justifications and comments), grouping statements into the rubric in Table  \ref{tab:RubricThemes} criteria.

\subsection{Evaluators Team}
One evaluator is a full professor of information systems and digital technologies. They research on human cognition, behaviour, and social interactions in software engineering. This evaluator brought extensive experience in QDA and research of human aspects of digitalisation and software development.

Another evaluator is a social scientist, an associate professor with a background in media and communications and marketing. They specialise in the social impacts of digital media and emerging technologies. Their research includes intimacy, online privacy, AI, humanoid robots, influencer marketing, and digital nomadism. They have extensive experience in QDA, specifically in thematic analysis.

One more evaluator holds a PhD in information systems and has over a decade of research and teaching experience. Their work spans interdisciplinary projects in digital technologies, healthcare, and mental health. They have expertise in design methods, technology adoption across cultural contexts, and connected health. They have experience in mixed methods and have performed thematic analysis extensively.

The last evaluator is a senior lecturer in computing science with expertise in software engineering, human–machine interaction, and sustainability. Their research addresses education, health, and well-being domains, where digitalisation reshapes traditional services and user expectations.

The evaluators represented a complementary mix of expertise in SE, qualitative research methods, human–technology interaction, healthcare, and socio-cultural studies. They were chosen to provide a broad yet rigorous perspective. They ensured that evaluations of LLM outputs considered methodological quality alongside human, organisational, and societal dimensions.

\subsection{Ethical Considerations}
We obtained informed consent from all interviewees to use their data. Transcripts were anonymised before analysis. We used the LLMs following their respective terms of service. Evaluators were blinded to the source of coded outputs to minimise bias in comparative assessment.



\section{Results} \label{sec:results}

This section presents the results of our evaluations of LLMs' performance in conducting Phases 2 through 5 of TA, following Braun and Clarke's six-phase framework. The section is organised in a chronological order, hence the code evaluations are presented first and then the theme evaluations.

\subsection{Evaluation Phase 2: Human vs LLM (1a)}

\begin{figure}[t]
    \centering
    \includegraphics[width=0.9\linewidth]{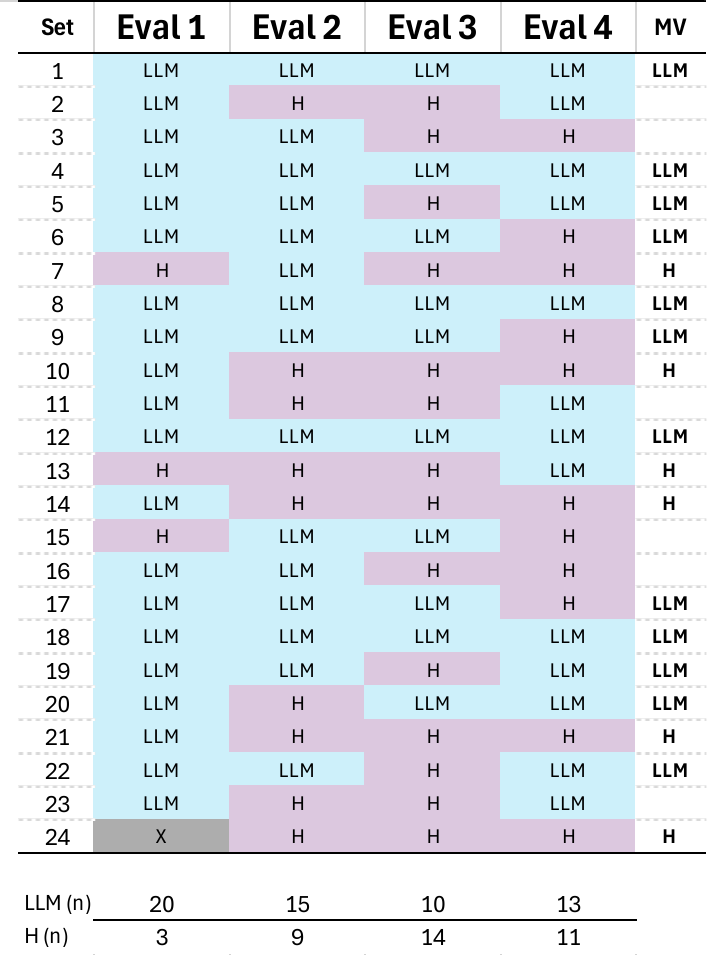}
    \caption{Evaluator preferences for human and LLM-generated codes across 24 interview segments. Each column corresponds to one of four evaluators (Eval 1–4), and each row represents one set of codes. Cells indicate whether the evaluator chose the human (H) or LLM (LLM) codes. Evaluator 1 declined to rate the final segment, judging that none of the codes aligned with the RQ. The majority voting (MV) column indicates majority preferences. H (n) and LLM (n) show how often each evaluator chose the H or LLM code.}
    \label{fig:first_results}
\end{figure}

As explained in Eval 1\ref{1a}, evaluators had to choose blindly between human and LLM codes. They preferred LLM-generated codes more often, 58 times out of 95 (since Evaluator 1 did not rate the final set of codes). The total rate was 61\%. Meanwhile, Human-generated codes were selected 37 times (39\%). Figure~\ref{fig:first_results} summarises their preferences across all segments. The choice for LLM codes was relatively consistent across evaluators. However, there was some variation; for example, Evaluator 3 selected human codes more often than others. Complete agreement among the four evaluators was shown only for LLM-generated codes five times.

\begin{figure}
    \centering
    \includegraphics[width=1\linewidth]{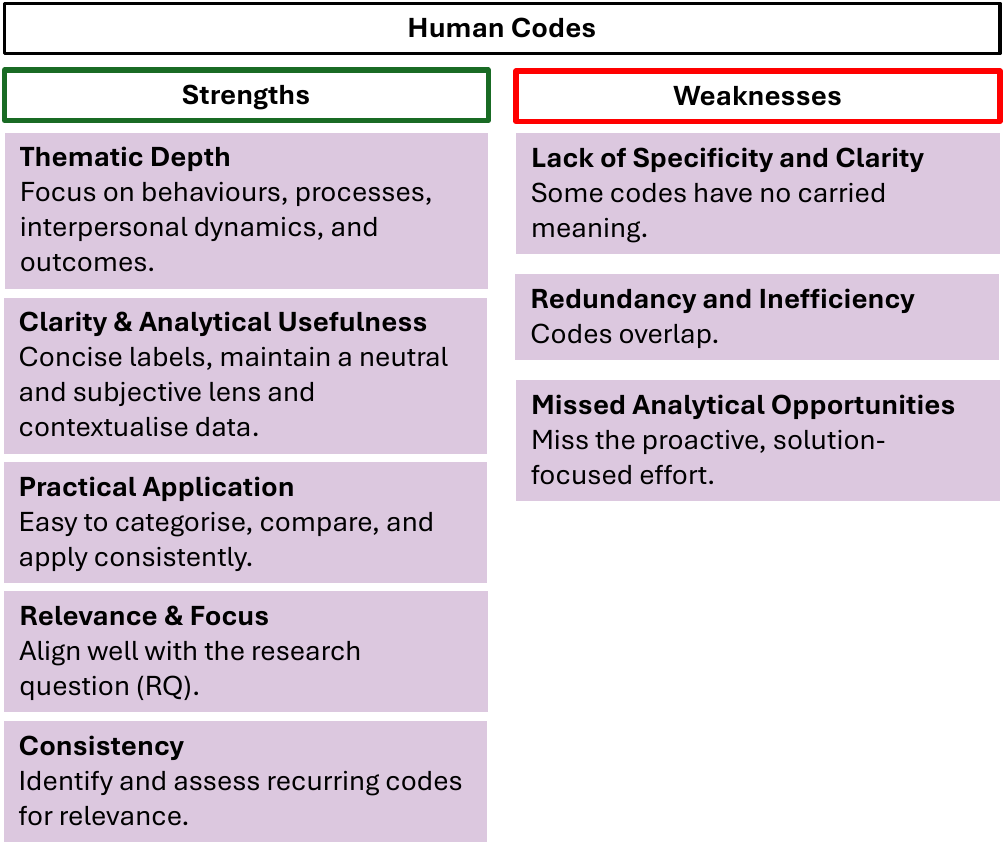}
    \caption{Summary of strengths and weaknesses identified by evaluators for human-generated codes, based on the Phase 2 (1a) evaluation.}
    \label{fig:human_codes_quali}
\end{figure}

\begin{figure}
    \centering
    \includegraphics[width=1\linewidth]{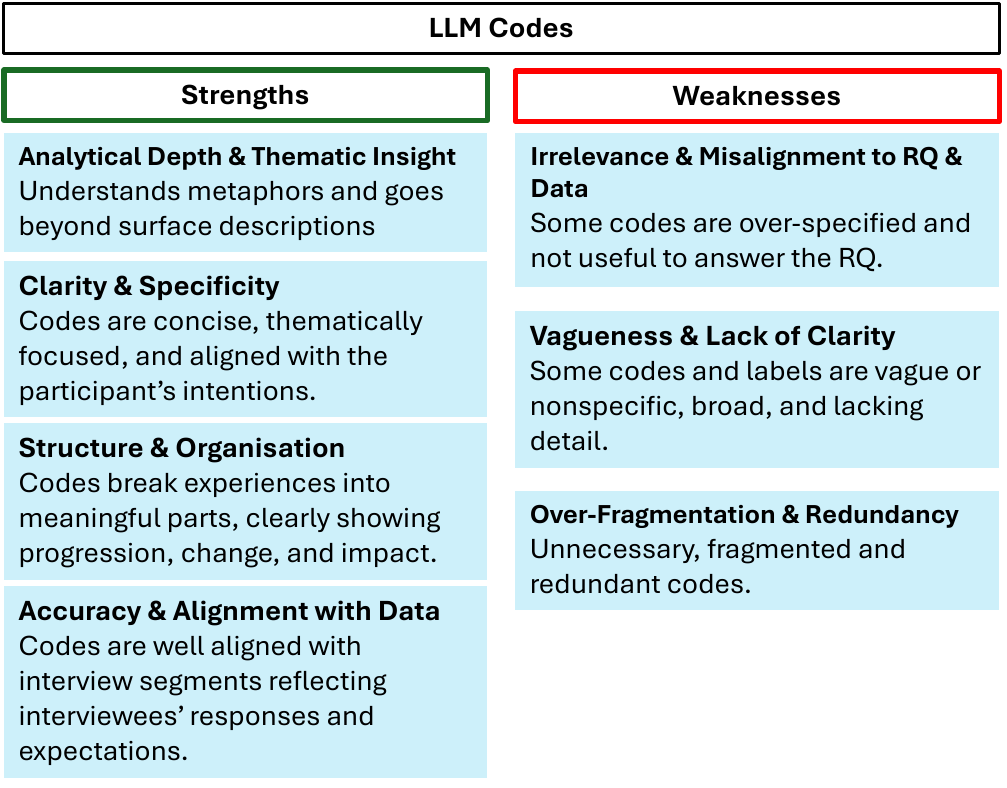}
    \caption{Summary of strengths and areas of opportunity identified by evaluators for LLM-generated codes, based on the Phase 2 (1a) evaluation.}
    \label{fig:llm_codes_quali}
\end{figure}

\bigbreak
The evaluation's \textbf{qualitative} part is summarised in Figure~\ref{fig:human_codes_quali} for human codes and Figure~\ref{fig:llm_codes_quali} for LLM-generated codes. The full table with all the comments is available in the online appendix \cite{martinez_montes_2025_17401526}.

Despite the lower overall selection rate, evaluators identified several strengths in \textbf{human-generated codes}. They noted that the codes demonstrated thematic depth. The codes captured behaviours, interpersonal dynamics, motivations, and outcomes without excessive fragmentation. Hence, codes offered breadth and clearer connections within the data. Evaluators also noted that the codes demonstrated clarity and analytical usefulness. Labels were concise and insightful, going beyond mere description to contextualise data and extract the core meaning of interviewee responses. Experts also mentioned codes' practical application, being easy to categorise, consistent in structure, and actionable in use. Similarly, they found the codes relevant and focused, since they aligned closely with the RQ and directly identified influences on well-being. Finally, evaluators mentioned that codes showed consistency across the sets.

At the same time, evaluators identified several areas of opportunity. A main concern was the lack of specificity and clarity. Some codes were deemed overly broad, generic, or descriptive, lacking nuance and depth, and occasionally failing to reflect participants' expressions accurately. Related to this, evaluators pointed to redundancy and inefficiency, with overlapping codes that reduced analytical sharpness. In addition, they noted that some codes missed analytical opportunities. This was particularly in capturing proactive or solution-oriented perspectives and clarifying how specific aspects of participant perceptions influenced outcomes.

\bigbreak
Regarding \textbf{LLM-generated codes}, Figure~\ref{fig:llm_codes_quali} presents evaluators' qualitative feedback. They observed substantial analytical depth and thematic insight. The codes extended beyond surface-level descriptions to reveal underlying factors, tensions, and contradictions. LLM outputs were noted for their ability to understand metaphors, identify inferred concepts, and link physical and mental aspects of well-being. Sometimes codes situated participant experiences within broader theoretical frameworks. This interpretive characteristic gave the codes a sense of "telling a story" rather than offering only descriptive labels. 

Another strength was clarity and specificity, reflected in the rubric scores for Clarity of meaning WA = 2.94 (Figure \ref{fig:rubric_results}). The codes frequently provided precise descriptions of important concepts, used concise action-oriented phrasing, and aligned well with participants’ intentions. The codes also displayed good structure and organisation. Codes broke experiences into meaningful parts, showing progression and change, and supporting comparison across diverse experiences. Finally, evaluators mentioned accuracy and alignment with the data. Many codes captured brief but significant excerpts and reflected participants' responses faithfully.

Evaluators also identified areas of opportunity. One recurring issue concerned relevance and alignment to the RQ and data. This improved in the second evaluation (WA = 3.04). Some codes were judged not directly helpful in answering the research question. Others were overly specific, focused too much on individual experiences rather than team or organisational perspectives, or misrepresented participant accounts. Evaluators also noted vagueness and lack of clarity. Some labels were broad, insufficiently action-oriented, or ambiguous in conveying causality and influence. Finally, problems of over-fragmentation and redundancy were reported. Some codes overlapped unnecessarily, while others fragmented experiences, reducing coherence.

\cristysbox{ \ta{Key finding:}Our results showed that, for Phase 1 (generating initial codes), LLMs can produce codes that are competitive with, and sometimes preferred over, codes produced by experienced human researchers.}

In this evaluation, evaluators chose LLM codes 61\% of the time. The reason for this high percentage can be due to the surface-level readability and polish of LLM outputs. Evaluators consistently described LLM codes as concise, specific, and well-formulated. This made codes appear clearer and easier to apply, even when they sometimes lacked deeper alignment with participants' intent. 
By contrast, while strongly relevant to the RQ and contextually sensitive, human codes were often longer, less standardised, and occasionally redundant. These characteristics can make human codes appear less precise in side-by-side comparisons, even if they carry greater reflexive or interpretive weight.

\subsection{Evaluation Phase 2: LLM Codes Rubric-Based Evaluation (1b)}

\begin{figure*}[h]
    \centering
    \includegraphics[width=0.9\linewidth]{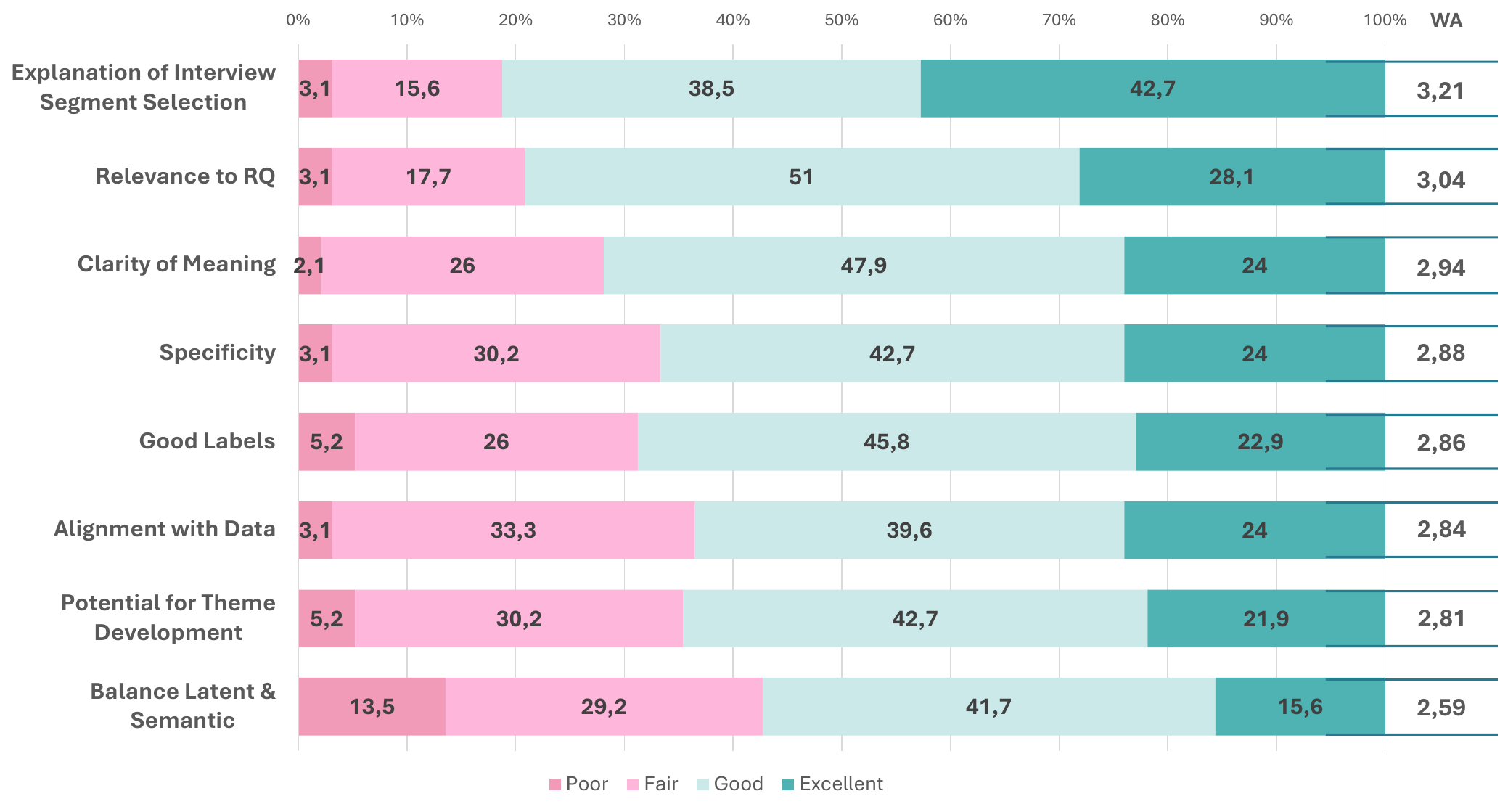}
    \caption{Rubric-based evaluation of LLM-generated codes across eight criteria (see Table \ref{tab:RubcodesEval}). Based on evaluations from four raters, bars indicate the percentage of ratings for each quality level (Poor, Fair, Good, Excellent). The weighted average (WA) per criterion is shown on the right.}
    \label{fig:rubric_results}
\end{figure*}

Figure \ref{fig:rubric_results} shows the ratings distribution across all evaluated code sets for the second part of the phase 2 evaluation (see section \ref{sec:phase2:b}). 
Overall, most criteria received a ``Good" rating. The better-rated dimensions were ``Explanation of Interview Segment Selection" (Excellent = 42.7, WA = 3.21) and ``Relevance to RQ" (Excellent = 27.1, WA = 3.04). This indicated the LLM's capabilities to justify segment selection and alignment with the RQ and the study's analytical goals. The lowest rated criterion was ``Balance between Latent and Semantic Codes" (Poor = 13.5, WA = 2.59). This showed that LLMs may have difficulties interpreting segments, which is expected in human-led qualitative analysis. Similarly, ``Potential for Theme Development" (WA = 2.91) was rated moderately, implying room for improvement in analytical precision and thematic extrapolation.

Similar to the previous evaluation, we asked evaluators to give feedback by explaining their assessment. Table \ref{tab:quali_rubric_codes} presents the analysis of evaluators' comments across eight criteria. The full table is in the online appendix \cite{martinez_montes_2025_17401526}. In this evaluation, experts focused mainly on the weak aspects of the codes; hence, the table is more populated on the negative side.

The two criteria with balanced comments were clarity of meaning and explanation of the interview segment selection. This finding aligned well with the quantitative evaluation in Figure~\ref{fig:rubric_results} (WA= 3.21). Regarding clarity of meaning, evaluators stressed the LLMs' ability to produce clear and coherent interpretations. LLMs performed well on this even when working with fragmented or ambiguous participant responses. However, they also commented on missed nuances, vague terminology, and occasional mismatches between the code's emphasis and the segment's actual intent.

More positively, the `explanation of interview segment selection' was consistently commented to be clear, coherent, and persuasive framing. The positive comments aligned well with the positive results from the rubric scores. At the same time, evaluators identified that strong justifications sometimes masked weak codes or introduced assumptions not fully grounded in the data.

\cristysbox{ \ta{Key finding:}Codes were good at clarity and relevance but weaker at nuance and latent interpretation.}

Our results indicate that LLMs can produce codes that meet several foundational standards of TA, particularly in terms of clarity and relevance. However, more abstract or interpretive dimensions, such as latent insight to create codes and interpretive nuance, seem more challenging to achieve.

\begin{table*}[h!]
\renewcommand{\arraystretch}{1.3}
\setlength{\tabcolsep}{8pt}
\centering
\caption{Qualitative feedback from evaluators on the performance of LLM-generated codes, grouped by rubric criterion. Positive comments start with + and negatives ones with -.}
\begin{tabular}{p{5cm} p{12cm}}
\hline
\textbf{Criteria} & \textbf{Comments} \\
\hline
\textbf{Clarity of Meaning} & 
+ Codes align closely with the expected interpretation \newline
+ LLM interpreted a fragmented and incoherent response surprisingly well \newline

- Missed some details and do not reflect suggestions \newline
- Inappropriate or unclear terminology \newline \\
\hline
\textbf{Relevance to Research Question} & 
+ Some codes are relevant to RQ \newline

- Some codes are not clearly linked to the RQ \newline
- Codes miss expressive or subjective content and nuanced segments \\
\hline
\textbf{Balance of Latent and Semantic} & 
+ Overall performance is fair in balancing latent and semantic content \newline
+ Most codes stay too close to the surface meaning of the text \newline

- There is untapped potential for deeper interpretation of participant meaning \\
\hline
\textbf{Specificity} & 
+ Descriptive codes match a descriptive response \newline

- Certain codes are too broad, redundant or unfocused and miss nuances \newline
- Codes are mostly descriptive and lack depth and detail \newline
- Codes are either too granular or lack granularity; the middle point is not there \\
\hline
\textbf{Potential for Theme Development} & 
- Missed opportunity to explore causes or implications \newline
- Codes may overlap in meaning; merging could improve thematic clarity \\
\hline
\textbf{Alignment with Data} & 
- Codes miss contextual references and key details from interviewees \newline
- Code and explanation ignore the question context \\
\hline
\textbf{Good Labels} & 
+ Code label summarises content effectively \newline
+ Avoids direct quotation from the text \newline

- Code label may narrow the meaning and reduce accuracy \newline
- Unclear if code reflects an observed event, an experience or a suggestion \newline
- Labels omit key details relevant for further analysis and are ambiguous or misleading \\
\hline
\textbf{Explanation of Interview Segment Selection} & 
+ Good explanation, is persuasive and compelling \newline
+ Explanation of code selection was strong and well-articulated \newline

- Risk of overestimating code quality due to strong explanation \newline
- Explanation includes assumptions not grounded in data, over-interprets participants’ words and does not consider the full context \\
\hline
\end{tabular}
\label{tab:quali_rubric_codes}
\end{table*}

\subsection{Evaluation Phase 3-5 Generating, Refining and Naming Themes (2a)}

\begin{figure*}[t]
    \centering
    \includegraphics[width=1\linewidth]{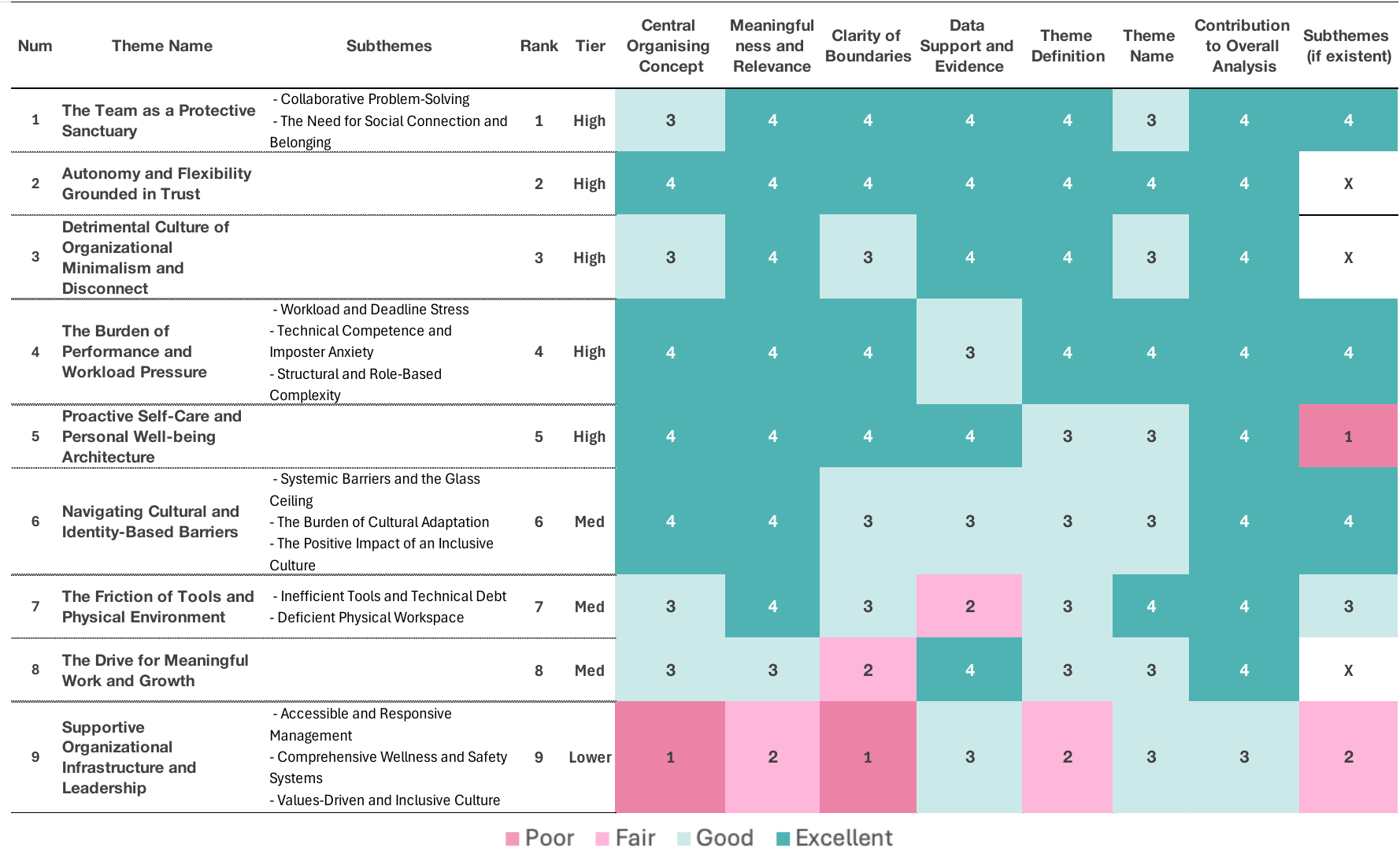}
    \caption{Rubric-based evaluation of nine themes produced by the best-performing LLM pipeline. The first five columns (``Num" to ``Tier") are LLM's outputs. The remaining columns show human evaluations based on the rubric in Table \ref{tab:RubricThemes}. Three evaluators rated each theme on a scale of 1 (poor) to 4 (excellent). The values shown are median scores across raters.}
    \label{fig:themes_eval}
\end{figure*}

Figure~\ref{fig:themes_eval} summarises the themes' evaluation results. The figure shows a decreasing tendency in the evaluations. It starts from a well-evaluated theme, ``The Team as a Protective Sanctuary", to a theme with low scores, ``Supportive Organisational Infrastructure and Leadership". The scores aligned well with the ranks and tiers proposed by the LLM. The first five themes were placed in the High tier, showing consistently strong performance across dimensions. 

Three themes were placed in the Medium tier. These typically showed solid organising concepts but were weak in clarity of boundaries. The lower score was in data support and evidence, since some codes that formed this theme were vague. 

The last theme, ``Supportive Organisational Infrastructure and Leadership", fell into the Lower tier due to limited analytical sharpness and inconsistent naming. It showed weak differentiation from other themes, unclear subtheme use, and insufficient clarity in tone and contribution to the overarching analysis.

These findings indicate that the selected LLM pipeline was capable of creating themes that meet many of the criteria of TA. While not all themes achieved equal strength, the best-performing ones have structural clarity, interpretive coherence, and alignment with participants' meaning.

On the weak side, experts commented on the apparent over-fragmentation of the theme structure. For example, Theme 9 could be merged with Theme 3 as a subtheme. Although Theme 9's topic is important, it overlaps with the central organising concept of Theme 3. The experts proposed creating a subtheme titled ``Organisational Infrastructure and Company Support" and including it under Theme 3.

Besides that suggestion of restructuring, the themes grouping seems complete in topics that help to answer the RQ. 

\cristysbox{ \ta{Key finding:}LLM themes were strong overall but sometimes fragmented; experts suggested merging overlaps.}

\section{Discussion} \label{sec:Discussion}

Following the current discussions around the inclusion of AI in QDA, we adapted Braun and Clarke's TA framework to be performed partially by LLMs. Our study addresses the call to design and tailor prompts for QDA \cite{de2024further}. In this section, we discuss the implications, limitations, benefits, methodological aspects, and recommendations to consider when using LLMs in QDA.

\subsection{LLMs as Analytical Assistants in SE Research}

Based on our results, we conclude that LLMs can perform several phases of TA. In particular, they are effective in initial coding and theme generation. Evaluators rated the LLMs’ outcome quality highly, specifically the codes, which were preferred over human-generated ones. These findings align with previous positive outcomes of conducting QDA using AI \cite{qiao2025thematic, drapal2023using, wiebe2025qualitative, nguyen2025chatgpt}. This thereby points to LLMs as viable analytical assistants in qualitative research. In contexts with large volumes of qualitative data or with a single researcher, LLMs can help reduce time and manual workload. They assist in these scenarios by offering codes and candidate themes that researchers can refine. However, this help comes with the need for caution. LLMs lack a critical and analytical perspective, which may compromise the method's rigour. Experts' reflexive supervision is necessary during the LLM's pre-coding and clustering. Similarly, during the interpretation phases, the model can generate convincing outputs that are not always grounded in the data. Because qualitative data analysis involves reflexive, complex, and continuous meaning-making \cite{manning2014making}, automating the entire process is not recommended. LLMs can surface patterns, candidate framings, and alternative readings, but they do not make meaning in the epistemological sense; that remains the role of reflexive human interpretation. Accordingly, LLM outputs should be positioned as scaffolds for researcher sense-making (e.g., prompts to compare, contest, or refine interpretations), not as substitutes for interpretive judgement or context stewardship.

\begin{figure}
    \centering
    \includegraphics[width=0.8\linewidth]{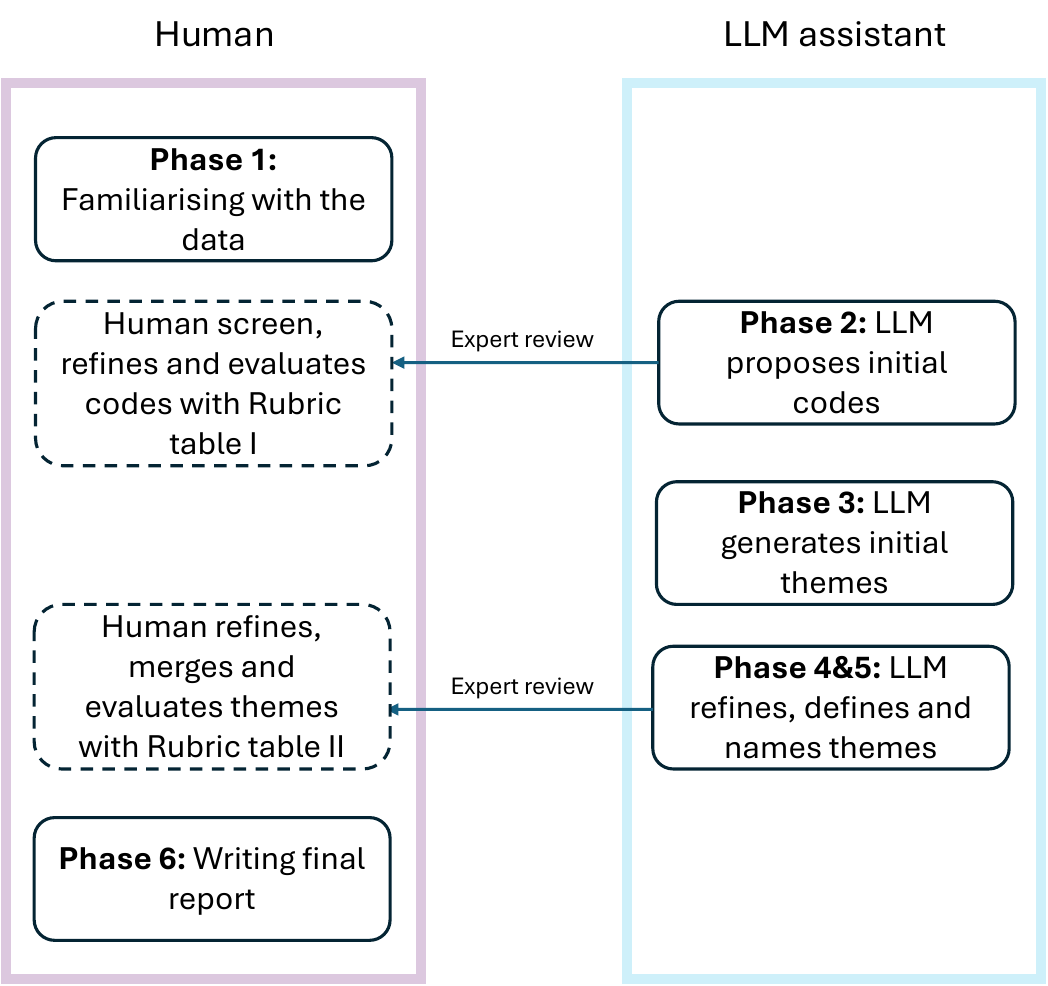}
    \caption{Proposal of implementation of LLM in TA. LLM contributes to TA Phases 2–5 as an assistant; the human leads Phases 1 and 6 and gates progression using rubric-based evaluations. Dashed boxes indicate areas that require human evaluation and refinement.}
    \label{fig:proposal}
\end{figure}

The following subsections elaborate more on the role and potential collaboration humans and LLMs can have when doing QDA together. Figure \ref{fig:proposal} presents LLM tasks as an analyst assistant. 

\begin{figure*}
    \centering
    \includegraphics[width=1\linewidth]{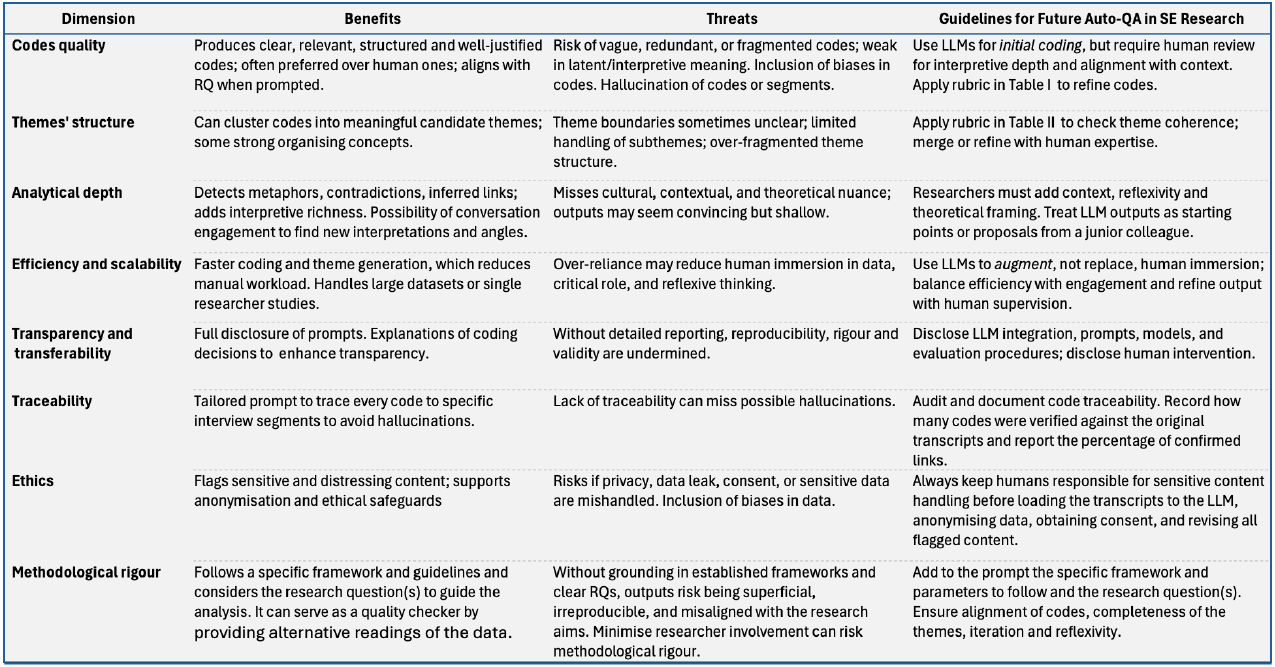}
    \caption{Reflection-based summary of the benefits, risks, and methodological guidelines for integrating LLMs in thematic analysis within SE research.}
    \label{fig:benefits_table}
\end{figure*}

Based on our discussion, we present in Figure \ref{fig:benefits_table} the benefits, risks and guidelines when integrating LLMs in QDA. It aims to support researchers in designing their studies more transparently and rigorously. It can be used as a reference point for evaluating and reporting LLM involvement in future empirical SE work.

\subsubsection{Human Oversight Remains Essential}

Our findings reaffirm that while LLMs can support qualitative analysis, they cannot replace human researchers. In line with prior work \cite{han2025can, nguyen2025chatgpt, bano2024large}, it is recommended to assign LLMs the role of assistant or collaborator instead of a substitute. One of the reasons is that we found that LLMs struggled to balance semantic and latent coding \cite{han2025can}. This matters because latent coding in SE research often requires connecting technical knowledge with implicit beliefs, values, or organisational dynamics. Such interpretations rely on human judgment and reflexivity \cite{braun2021thematic}. Human oversight is essential for ensuring nuanced and valid interpretations\cite{nguyen2025chatgpt}.

Furthermore, we found that LLMs risk oversimplifying or misinterpreting without feedback and supervision, consistent with \cite{han2025can}. Essential aspects such as tone, intention, and power dynamics remain accessible only through human engagement in the research process. As we advocate for a collaboration with LLMs in QDA, we also emphasise that researchers must remain knowledgeable about their dataset. They have to be able to guide the LLM, and vigilant about bias and alignment with research goals \cite{christou2023use}. Transparency about the involvement of LLMs is also a key responsibility.

\subsection{Strengths and Limitations of LLM Outputs in Engineering Contexts}

The strengths of LLM-generated codes (see Figure \ref{fig:llm_codes_quali}) include Analytical Depth \& Thematic Insight, and Structure \& Organisation, which are important characteristics when doing QDA. Evaluators even commented on the capacity to understand metaphors and identify inferred concepts. This is notable given the topic in our dataset. 

Moreover, the LLM could clearly justify codes creation in phase 1. During the evaluation, this criterion received the highest score in ``Excellent" (see Figure \ref{fig:rubric_results}, Explanation of Interview Segment Selection WA = 3.21). It could also explain why a particular quote or excerpt was used in a theme. This results from tailoring the prompt with a theoretical position to simulate ``choosing" segments based on their research purpose. However, it is important to remember that this is only a mimic, not an alignment with epistemological expectations.

Relevance to Research Question WA = 3.04 was also among the highest scores. The LLM could, in general, align codes with the RQ and avoid tangential interpretations. We included the RQ in the prompt along with specific guidelines of what type of codes we expected. This guided the LLM in prioritising analytical decisions for the study's goal. With a non-tailored prompt, the LLM might fail to choose relevant codes and output surface-level or not well-aligned segments. It is important to note that evaluators commented that some codes did not appear to be fully aligned with the RQ. This was in phase 2, where the first coding round was done. Braun and Clarke clarify that codes and even themes are not final until the end of the analysis. Having tangential codes that could potentially be added later to the analysis does not translate into a problem. In this case, it reinforces the need for researchers to review, refine and guide the iterative stages of analysis.

However, there are limitations regarding Vagueness and lack of Clarity and Redundancy. Especially in Phases 4–5, themes occasionally lacked clear boundaries. LLMs did not create subthemes when there was room for the,m and appeared too fragmented in their theme structure. These shortcomings can reduce the coherence of the analysis and fail to capture nuances in the data. As result, this can compromise the narrative to answer the research questions.

Furthermore, the lowest rubric score was Balance of Latent and Semantic, WA = 2.59. It is essential to have both types of codes in thematic analysis to create a wholesome analysis. LLM did a good job at the semantic level. However, creating latent codes requires theoretical reading, ideological critique, and cultural and contextual sensitivity, which the LLM fails to mimic. This result shows the limitations and risks of relying only on LLMs to create codes. It once more stresses the importance of researchers in identifying deeper meanings, questioning assumptions, and situating findings within broader theoretical and socio-cultural frameworks.

One more factor to consider is the clarity and formulation quality in LLM's outputs. LLM outputs are often linguistically polished and consistent, which can influence judgments of quality even when interpretive depth is limited. Thus, evaluator preference does not necessarily equate to epistemic adequacy. This situation might be more critical when qualitative data comes from a context where meaning is expressed through colloquial, non-standard, or culturally embedded language.

\subsection{Strategies to Ensure Methodological Rigour}
We implemented the following strategies based on Christou's \cite{christou2023use} suggestion of the need to ensure accuracy and credibility in all AI-generated content by cross-referencing it. 

\paragraph{ } The dataset we used and the prompt had cues that helped us \textbf{trace} back the source data (each segment for each code per interview). By instrumenting and cross-referencing the prompt, we \textbf{mitigated hallucinations, lack of transparency, inconsistency and increased trust and auditability}. To strengthen this even more, we required the LLM to work segment by segment and to generate codes anchored in specific data excerpts to increase data fidelity. 

\paragraph{ } We also prompted the LLMs to write an explanation of the selected codes and the reason each code helped answer the RQ. With this, we aimed to improve \textbf{transparency} in each step of the process. We also enable more precise alignment between the data, the generated codes, and the main research question.

\paragraph{ } To ensure \textbf{methodological accuracy}, we instructed the LLM specifically with the Braun and Clarke Thematic Analysis approach. We included the definition and examples of codes and specifications of the type of coding, inductive, in this case. To complement this, we tailored our prompt to have a specific research purpose by giving it a research question.

\paragraph{ } Additionally, we requested the LLM to flag sensitive content in the interview segments to ensure the \textbf{ethical handling} of potentially distressing material and support AI's responsible use in qualitative analysis.

Combining these strategies, we aimed to provide a structured and transparent method to use LLMs in TA. We included strategies to ensure methodological rigour and to address AI's common challenges \cite{bano2024large}. These relate to the credibility, ethical integrity, and trustworthiness of the analysis process. Process rigour is maintained by positioning the LLM as an analytical assistant while the researcher ensures reflexivity, iterative engagement, and interpretive decision-making.

\subsection{Implications for Empirical SE Methods}

As the integration of LLMs in qualitative research, particularly in QDA, appears inevitable \cite{wiebe2025qualitative}, it is essential to define and assess their role and limitations. We must also examine the methodological implications of their use in SE research and beyond. We propose to have them as analytical assistants in qualitative research. This implies rethinking current strategies for ensuring quality and trustworthiness in qualitative research. Such quality features are credibility, transferability, dependability and confirmability \cite{ahmed2024pillars, christou2023use}.

\begin{itemize}
    \item Credibility: Researchers need to familiarise themselves with their data to give feedback to the LLM throughout the QDA process. Regarding reflexivity, it must expand to critically assess the AI's biases, limitations, and how its outputs influence interpretations. Researchers need to reflect not only on their subjectivity but also on how the AI shapes the analytic process and outcomes.
    \item Transferability: Clear reporting standards should now include details about the research context, model specification, and human involvement. They should also describe the prompt structure and how the AI output was integrated into the analysis.
    \item Dependability: The methodological documentation should explicitly detail the AI components, preprocessing steps, and how researchers validated or modified AI outputs. Audit trails must register AI interactions, results and decision-making by the researcher based on them. A change tracker on codes or themes changes can aid with this marker.
    \item Confirmability: For peer debriefing, discussions about the AI's role and input need to be included. Building on the previous point, the influence of AI in data interpretation needs to be questioned. Member-checking practices also need adaptation. If the study's participants are asked to review AI-generated summaries or themes, researchers must clearly explain the role of AI in the process. Additionally, it is necessary to consider participants' views on AI's interpretative role. Finally, researchers' reflexive journals should include insights on AI-related challenges and decisions. They should also acknowledge how AI's presence shaped the researcher's thought process and interpretations.
\end{itemize}

Integrating LLMs as analytical assistants in SE can help with specific field challenges. Such challenges are managing massive, technically rich, and inherently socio-technical data. Since SE qualitative data often combines technical artefacts with human-centric sources. LLMs can help by triangulating insights more efficiently and effectively with large amounts of data. In the requirements engineering area, for example, LLMs can easily and consistently trace data. This will allow human researchers to focus on higher-level interpretative work. Additionally, they can maintain consistency across long or multi-researcher projects, aid in reducing mental exhaustion, and assist the researcher in identifying biases. Prompts can be tailored to try different QDA, which gives the researcher a perspective on data analysis decisions. 

Finally, it is equally important to address ethical and privacy aspects. In this study, we handled all data under informed consent and anonymisation protocols. LLMs processed no personal identifiers. All models were accessed under institutional terms of service, ensuring compliance with privacy standards. We, researchers, were responsible for ethical oversight and for verifying sensitive segments flagged by the models. However, more researcher is needed to tailor formal frameworks for responsible AI-assisted qualitative analysis. This is especially important in cases where qualitative data contains sensitive or personal information.

\subsection{Threats to Validity}
We took several measures to strengthen the validity of our findings across the four standard categories: internal, external, construct, and conclusion validity. We elaborate on them in the following paragraphs.

\textbf{Internal validity:} Our dataset included a ground truth of pre-coded interviews by two experienced researchers. Additionally, we employed blinded comparative evaluation during Phase 2. This ensured that evaluators did not know the code generators. As a result, it helped isolate the effect of code quality from potential biases related to authorship. We also iteratively refined prompts and evaluation procedures to reduce confounding variables related to prompt phrasing or interpretation. Having expert evaluators with prior qualitative research experience increased consistency in applying evaluation criteria, reducing potential noise in the assessment process.

\textbf{Construct validity:} Our prompts were lengthy, as we included a significant amount of information and quality checks. This might make them complex and harder to use. To mitigate this, we structured the prompts clearly, provided step-by-step instructions, and tested them iteratively to ensure usability and effectiveness.

Rubrics were grounded in Braun and Clarke’s Reflexive TA framework, ensuring alignment with TA rigour standards. Additionally, we prompted the LLMs to explain their coding decisions and how they related to the RQ to ensure that outputs reflected more than superficial content features.

\textbf{Conclusion validity:} We triangulated evaluators' feedback by having experts from social sciences and SE. Furthermore, we collected quantitative and qualitative data from their evaluations to make more robust inferences about LLM performance. This allowed us to cross-check results across different evaluative lenses. Furthermore, we evaluated five complete LLM pipelines, rather than relying on isolated examples. We selected the best-performing one through a systematic comparison, which strengthened our conclusions.

\textbf{External validity:} The prompt was tailored in a way that can be used with any qualitative dataset. We acknowledge that LLM performance may vary across topics and datasets, and future work should explore broader generalisability. Additionally, we only tested code quality within a specific topic and a homogeneous population. It is necessary to account for differences in colloquial language, culturally embedded experiences, and diverse types of populations.

\section{Conclusion}

This study provided one of the first systematic, rubric-based evaluations of  LLMs as analytical assistants in qualitative SE research. Our results showed that LLMs can produce analytically applicable and often well-structured codes and themes. Human evaluators preferred LLM-generated codes in most cases (61\%), confirming their potential to augment human interpretation. 

In this study, we offer a documented, reproducible framework including:
\begin{itemize}
    \item Complete prompts for use in LLMs. 
    \item Tailored rubrics based on Braun and Clarke's TA to evaluate the quality of codes and themes. 
    \item Result-based guidelines to integrate LLMs into QDA.
\end{itemize}

We particularly stress the methodological grounding of our study to ensure rigour and trustworthiness. We conclude that LLMs can assist in QDA in SE when used as collaborators. This is effective as long as they are embedded within well-defined methodological and ethical boundaries. To examine the process integrity of using LLMs as analytical assistants in thematic analysis, future work should conduct parallel analyses on new data, comparing a fully human analysis with an LLM-assisted one.

\section{Authors’ Contributions}
C.M.M. and R.F. conceived and designed the study. R.F. implemented and ran the LLM prompts, provided feedback, and contributed to shaping the study design. C.M.M. performed the data analysis, integrated the evaluation results, and prepared the manuscript drafts. Both authors discussed the findings and contributed to the interpretation of the results. 

The evaluators, C.M.M., S.P., S.O., and D.G., conducted the blind evaluations of the codes and themes and revised the final paper version, but did not participate in the study design, or prompt development.


\bibliographystyle{ieeetr}
\bibliography{bib}
\vspace{12pt}
\color{red}

\end{document}